\documentclass{aa}

\usepackage[varg]{txfonts}

\usepackage[bookmarks=true,
            bookmarksopen=true,
            pdfstartview = FitV,
            pdfpagelayout = SinglePage,
            colorlinks = true,
            linkcolor=black,
            urlcolor = blue,
            pdfborder = {0 0 0}]{hyperref}

\begin{document} 

   \title{Probing the face-on disc-corona system of the bare AGN Mrk\,110 from UV to hard X-rays: a moderate changing-state AGN?}
\titlerunning{Probing the face-on disc-corona system of Mrk\,110}
   \author{D.\ Porquet \inst{1}  
          \and
          S.\ Hagen\inst{2}
          \and
          N.\ Grosso\inst{1}
          \and
            A.\ Lobban\inst{3}
          \and
            J.~N.\ Reeves\inst{4,5}
          \and
          V.\ Braito\inst{5,4,6}
           \and            
           C.\ Done\inst{2}
         }
   \institute{Aix Marseille Univ, CNRS, CNES, LAM, Marseille, France 
              \email{delphine.porquet@lam.fr}
              \and{Centre for Extragalactic Astronomy, Department of Physics, Durham University, South Road, DH1 3LE, UK}
         \and European Space Agency (ESA), European Space Astronomy Centre (ESAC), E-28691 Villanueva de la Cañada, Madrid, Spain
        \and  Department of Physics, Institute for Astrophysics and Computational
Sciences, The Catholic University of America, Washington, DC 20064, USA
          \and INAF, Osservatorio Astronomico di Brera, Via Bianchi 46 I-23807 Merate
 (LC), Italy
           \and Dipartimento di Fisica, Universitá di Trento, Via Sommarive 14, Trento 38123, Italy.
              }
   \date{Received , 2023; accepted , 2023}

 
  \abstract
  {The X-ray broadband spectra of the bare AGN Mrk\,110, obtained by simultaneous {\sl XMM-Newton} and \linebreak {\sl NuSTAR}  observations performed on November 2019 and April 2020, are characterised by the presence of a prominent and absorption-free smooth soft X-ray excess, moderately broad \ion{O}{vii} and Fe\,K$\alpha$ emission lines, and a lack of a strong Compton hump. The disc-corona system is almost viewed face-on as inferred from the \ion{O}{vii} accretion disc lines. While relativistic reflection as the sole emission is ruled out, a simplified combination of soft and hard Comptonisation (using {\sc comptt}) from a warm and a hot coronae, plus  mild relativistic disc reflection (occuring at a few 10s $R_{\rm g}$) reproduces the data very well.}
   {We aim to confirm the physical origin of the soft X-ray excess of Mrk\,110 and to determine its disc-corona system properties from its energetics using two new sophisticated models: {\sc reXcor} and {\sc relagn}, respectively.}
   {We apply these models to the 0.3--79\,keV X-ray broadband spectra and to the spectral energy distribution (SED) from UV to hard X-rays, respectively.}
 {At both epochs, the inferred high-values of the warm-corona heating from the X-ray broadband spectral analysis using {\sc reXcor}  confirm that the soft X-ray excess of Mrk\,110 originates mainly from a warm corona rather than relativistic reflection.
  The intrinsic best-fit SED determined at both epochs using {\sc relagn} show a high X-ray contribution relative to the UV
  and are very well reproduced by a warm and hot coronae plus mild relativistic reflection. The outer radii of the hot and warm coronae are located at a few 10s and $\sim$100 $R_{\rm g}$, respectively. Moreover, combining the inferred low Eddington ratio ($\sim$ a few percent) from this work, and previous multi-wavelength spectral and timing studies  suggests that Mrk\,110 could be classified as a moderate changing-state AGN.}
 {Our analysis confirms the existence of a warm corona as a significant contribution to the soft X-ray excess and UV emission in Mrk\,110, adding to growing evidence that AGN accretion deviates from standard disc theory. This strengthens the importance of long-term multi-wavelength monitoring on both single targets and large AGN surveys to reveal the real nature of disc-corona system in AGN.}

 \keywords{X-rays: individuals: Mrk\,110 -- Galaxies: active --
     (Galaxies:) quasars: general -- Radiation mechanism: general -- Accretion, accretion
     discs.}
   \maketitle
%

\section{Introduction}\label{sec:Introduction}

In the standard picture, the emission of an active galactic nucleus
(AGN) stems from an accretion disc around a \linebreak supermassive black hole (SMBH) with
masses spanning from a few million to billions of solar masses.
X-ray spectra allow us to probe the geometry and the main physical process(es) at work
in the inner part of the disc-corona system; for example:
relativistic reflection resulting from the illumination of the standard accretion disc by a hot corona
and/or Comptonisation of seed photons from the accretion disc by a warm-hot corona.
\citep[e.g.,][]{Magdziarz98,Porquet04a,Crummy06,Bianchi09,Fabian12,Done12,Porquet18,Petrucci18,Gliozzi20,Waddell20}.

A large fraction of AGN exhibit warm absorbers along the line-of-sight
\cite[e.g.,][]{Porquet04a,Piconcelli05,Tombesi13,Laha14a} which can severely bias the X-ray data analysis.
Therefore, bare AGN that show no (or very weak) X-ray warm absorbers are the best targets for directly investigating the process(es) at work in disc-corona systems around SMBH.
Very high signal-to-noise ratio data from simultaneous X-ray broadband observations of bright bare AGN with {\sl XMM-Newton} and \linebreak {\sl NuSTAR} offer us the possibility to determine the dominant physical process(es) at work, as performed, for instance, for the broad-line Seyfert\,1 (BLS1) \object{Ark\,120} \citep{Matt14,Porquet18,Porquet19} and the narrow-line Seyfert\,1 (NLS1) \object{TON\,S180}  \citep{Matzeu20}. These deep X-ray observations allow us to rule out, for these two AGN, relativistic reflection from a constant-density, flat, standard accretion disc as the sole emission process. Indeed, this emission process cannot, whatever the hot-corona geometry is, simultaneously reproduce the soft X-ray excess, the broad Fe K$\alpha$ complex, and the hard X-ray shape. Instead, the X-ray broadband spectra of these two AGN are very well reproduced by a combination of soft and hard Comptonisation from a warm and hot corona, respectively, plus mild relativistic disc reflection. \\

\object{Mrk 110} (also known as PG\,0921+525) is a bright, type\,1 Seyfert galaxy in the local Universe (z=0.035291),
with a bolometric luminosity of about 5$\times$10$^{44}$\,erg\,s$^{-1}$ \citep{Woo02}.
This source is radio-quiet but not radio-silent \citep{XuC99,MillerP93,Kukula98,Jarvela22,Panessa22,Chen22}.
From optical observations, its type E host galaxy displays a disturbed morphology
 with a significant tidal tail which could suggest a past merger or tidal
interaction with only one apparent nucleus \citep{Adams77,Wehinger77,Hutchings88,MacKenty90,Bischoff99}.
Moreover, both its optical continuum and broad lines are strongly variable on days-to-year time-scales \citep{Peterson84,Peterson98,Bischoff99,Kollatschny01,Homan23}.
In X-rays, Mrk\,110 exhibits flux variations by a factor of up to $\sim$4--5 on yearly time-scales, as observed by  long-term observations with the {\sl Rossi X-ray Timing Explorer} \citep[{\sl RXTE}; e.g,][]{Markowitz04,Weng20}.

This source is frequently classified as a NLS1 when
 only considering the measurements of a relatively narrow optical H$\beta$ emission line with a full width at half maximum (FWHM) of $\sim$1700--2500 km\,s$^{-1}$, emitted by its broad-line region \citep[BLR;][]{Osterbrock77,Peterson84,Crenshaw86,Bischoff99,Vestergaard02,Grupe04a}.
 However, Mrk\,110 displays much broader and more redshifted optical BLR components from Balmer lines (H$\alpha$, H$\beta$, H$\gamma$), \ion{He}{i}$\lambda\lambda$5876,6678 and \ion{He}{ii}$\lambda$4686 lines; (FWHM$\sim$5000--6000\,km\,s$^{-1}$; \citealt{Bischoff99,Veron-cetty07}),
as well as an unusually large [\ion{O}{iii}]$\lambda$5007/H$\beta$
ratio and very weak \ion{Fe}{ii} emission \citep{Boroson92,Grupe04a,Veron-cetty07}.
 Therefore, these properties are more consistent with a BLS1.
This is strengthened by the X-ray timing and spectral characteristics of Mrk\,110
which are similar to those found for moderate accretion-rate BLS1s   \citep{Porquet04a,Boller07,Piconcelli05,Zhou10b,Ponti12,Gliozzi20,Porquet21}.

Mrk\,110 hosts a SMBH with a well-constrained mass value
of 1.4$\pm$0.3$\times$10$^{8}$\,M$_{\odot}$, measured
from the detection of gravitationally-redshifted emission in the variable
component of all of the broad optical lines \citep{Kollatschny03,Liu17}.
Black hole (BH) mass values inferred for Mrk\,110 from different, independent methods
 are in agreement with this value: optical
spectro-polarimetric observations (log($M_{BH}$/M$_{\odot}$)=8.32$\pm$0.21; \citealt{Afanasiev19}), X-ray
excess variance (log($M_{\rm BH}$/M$_{\odot}$)=8.03$^{+0.40}_{-0.30}$; \citealt{Ponti12}), and X-ray scaling methods (log($M_{\rm BH}$/M$_{\odot}$)=8.2--8.5; \citealt{Williams18}).
A much lower mass value using the virial method ($\sim$2$\times$10$^{7}$\,M$_{\odot}$) has also been measured \citep{Peterson04}.
However, this latter method strongly depends on the inclinaton of the disc-like BLR system,
contrary to the gravitational method.
Indeed, as reported by \cite{Liu17}, the very large discrepancy between gravitational
and virial masses can be explained by an adapted $f_{FWHM}$ value of 8--16 for the virial measurement
that depends on the disc-like BLR inclination. The commonly used value of $f_{\rm FWHM}$$\sim$1
is relevant for an inclination angle of about 30 degrees.
This would imply a disc-like BLR system viewed almost face-on in Mrk\,110 \citep{Kollatschny03,Decarli08,Liu17}.\\

\begin{figure}[t!]
\begin{tabular}{c}
\includegraphics[width=0.6\columnwidth,angle=-90]{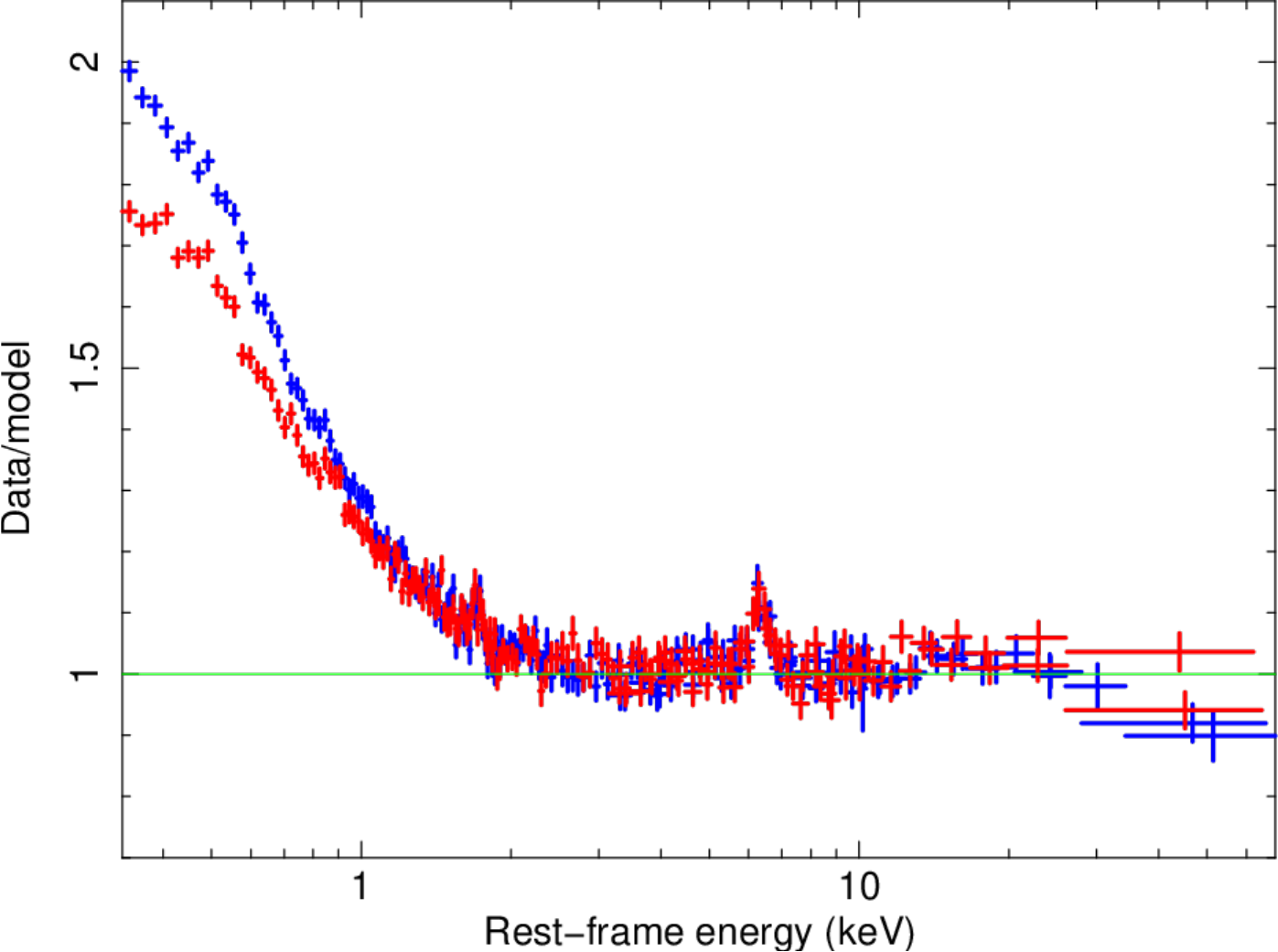} \\
\end{tabular}
\caption{Data-to-model ratio of the two simultaneous 2019 (blue) and 2020 (red) {\sl XMM-Newton}-pn
and {\sl NuSTAR} spectra of Mrk\,110 fit with a power-law model (with
Galactic absorption) over the 3--5 and 7--10 keV energy ranges and then extrapolated over the whole energy range. }
\label{fig:extrapolation}
\end{figure}

Mrk\,110 was observed twice, simultaneously by {\sl XMM-Newton} and {\sl NuSTAR}, on November 16--17 2019 and April 5--6 2020 \citep{Porquet21,Reeves21b}. The X-ray flux was a factor of $\sim$1.2 higher in the first observation compared to the second.
From the long-term X-ray behaviour of Mrk\,110 as observed by {\sl RXTE}
\citep{Weng20}, the observations appear to be consistent with a moderately high state of the source.
As shown in Fig.~\ref{fig:extrapolation}, the two X-ray broadband spectra ({\sl XMM-Newton} pn and {\sl NuSTAR}) are characterised by the presence of a prominent and absorption-free smooth soft X-ray excess, a weak Fe\,K$\alpha$ line, and a lack of a strong Compton hump. \cite{Porquet21} find for Mrk\,110 (as for Ark\,120 and TON\,S180) that relativistic reflection alone is not able to reproduce the soft X-ray excess and the hard X-ray spectral shape.
Instead, a combination of soft (using {\sc comptt}) and hard Comptonisation from a warm and hot coronae, respectively, plus mild relativistic disc reflection is needed to reproduce the broadband X-ray continuum. Its inferred warm corona temperature, $kT_{\rm warm}$$\sim$0.3\,keV, is similar to the values found in other sub-Eddington AGN \citep[e.g.][]{Gierlinski04,Porquet04a,Bianchi09,Petrucci18}. Its hot corona temperature, $kT_{\rm hot}$$\sim$30 keV \citep{Porquet21,Pal23}, is in the lower range of the average value measured from large samples of type-I radio-quiet AGN \citep[e.g,][]{Middei19,Panagiotou20,Akylas21,Kamraj22,Kang22}.

The presence of a broad \ion{O}{vii} soft X-ray emission line,
first identified by \cite{Boller07}, was confirmed
by the spectral analysis of the {\sl XMM-Newton} resolution grating spectrometer (RGS) data obtained between 2004 and 2020 \citep{Reeves21b}.
The \ion{O}{vii} line flux varies significantly
with the soft X-ray continuum flux level, being brightest when
the continuum flux is highest \citep{Reeves21b}, similar to the reported behaviour
of the optical \ion{He}{ii} line \citep{Veron-cetty07}.
This \ion{O}{vii} line originates from the accretion disc at a distance of a few tens of
gravitational radii ($R_{\rm g}$$\equiv$$GM_{\rm BH}/c^{2}$). The inclination angle of the accretion disc has been
well constrained (9.9$^{+1.0}_{-1.4}$ degrees) from the spectral analysis of these
\ion{O}{vii} lines \citep{Reeves21b}. This is consistent with an almost
face-on view of the disc-corona system, as also inferred for the disc-like BLR \citep{Bian02,Kollatschny03,Liu17}.
In all the RGS spectra, no significant intrinsic X-ray warm absorption
is present, with an upper limit for its column density of only 2.6$\times$10$^{20}$\,cm$^{-2}$,
demonstrating that Mrk 110 is a {\it bare} AGN, irrespective of its flux level \citep{Reeves21b}.
Mrk\,110 is, therefore, an excellent target to probe its disc-corona system, which is viewed almost face-on and without any significant neutral or warm gas in its line-of-sight. \\

\begin{table*}[!t]
\caption{Observation log of the two simultaneous {\sl XMM-Newton} and {\sl NuSTAR} datasets.}
\begin{tabular}{c@{\hspace{10pt}}c@{\hspace{10pt}}l@{\hspace{10pt}}l@{\hspace{10pt}}c@{\hspace{10pt}}c@{\hspace{10pt}}c}
\hline \hline
Mission & Obs.\,ID & Obs.\,start (UTC) & Obs.\,end (UTC) & Exp. (ks)$^{(a)}$ & count\,s$^{-1}$$^{(b)}$ \\
\hline 
\hline
{\sl NuSTAR}     & 60502022002 & 2019 November 16 -- 03:31:09 & 2019 November 18 -- 00:56:09 & 86.8,86.2 & 0.64,0.62  \\
{\sl XMM-Newton} &  0852590101 & 2019 November 17 -- 09:02:57 &  2019 November 17 -- 21:24:37 & 29.9 & 20.7\\
{\sl NuSTAR}     & 60502022004 & 2020 April 5 -- 14:26:09 & 2020 April 7 -- 13:26:09& 88.7,87.8 &  0.55,0.53 \\
{\sl XMM-Newton} & 0852590201 & 2020 April 6 -- 22:26:50 & 2020 April 7 -- 11:55:10 & 32.7 & 16.2 \\
\hline
\hline
\end{tabular}
\label{tab:log}
\flushleft
\small{\textit{Notes.} $^{(a)}$ Net exposure time (livetime corrected from any background flaring period) for {\sl XMM-Newton}-pn and  {\sl NuSTAR} (FPMA, FPMB). $^{(b)}$ Net source count rate over 0.3--10 keV for {\sl XMM-Newton}-pn and over 3--79\,keV for {\sl NuSTAR} (FPMA,FPMB).}
\end{table*}

Following the first X-ray broadband spectral analysis of Mrk\,110 using the two simultaneous 2019 and 2020 {\sl XMM-Newton} and {\sl NuSTAR} observations \citep{Porquet21}, here we aim to probe the physical properties of its disc-corona system properties based on sophisticated models that were recently released to the community. As a first step, we use the recent X-ray {\sc reXcor} model \citep{Xiang22} to estimate the warm-corona heating fraction and then to confirm whether or not the soft X-ray excess can physically (mainly or at least partly) originate from a warm corona, as previously inferred using a more simplified modelling \citep{Porquet21}.
We then perform an in-depth SED analysis from UV to hard X-rays using the {\sc relagn} model \citep{Hagen23b} -- adding a relativistic reflection model ({\sc relxillcp}; \citealt{Dauser13}) -- to infer the accretion rate and the physical properties of both the warm and hot coronae.
In Sect.~\ref{sec:obs}, the data reduction and the analysis methods of the dataset are presented.
The X-ray broadband analysis using the {\sc reXcor} model is performed in Sect.~\ref{sec:rexcor}.
In Sect.~\ref{sec:SED}, the SED (from UV to hard X-rays) analysis using the {\sc relagn} model complemented by relativistic reflection is reported.
The main results are discussed in Sect.~\ref{sec:discussion} and the conclusions are reported in Sect.~\ref{sec:conclusion}.\\

\section{Observations, data reduction, and analysis method}\label{sec:obs}
\subsection{{\sl XMM-Newton} and {\sl NuSTAR} data reduction}

The log of the simultaneous {\sl XMM-Newton} and {\sl NuSTAR} observations of Mrk\,110
({\sl NuSTAR} cycle-5; PI: D.\ Porquet) used in this work is reported in Table~\ref{tab:log}.
The {\sl XMM-Newton}/EPIC (European Photon Imaging Camera) event files were
reprocessed with the Science Analysis System (SAS; version 20.0.0), applying the
latest calibration available on November 21, 2022. Due to the high source
brightness, the EPIC-pn instrument was operated in the Small Window mode. 
 We note that only the EPIC-pn \citep{Struder01} 
data were used (selecting the event patterns 0--4, that is to say,
single and double events) since they do not suffer from pile-up (contrary to the EPIC-MOS data; \citealt{Turner01})
 and have a much better sensitivity above $\sim$6\,keV. 
The pn spectra were extracted from a circular region centred on Mrk\,110, with a radius of
35${\arcsec}$ to avoid the edge of the chip. The background
spectra were extracted from a rectangular region in the lower part of
the small window that contains no (or negligible) source photons.  
 The total net exposure times, obtained after the correction for dead time and background flaring, are reported
 in Table~\ref{tab:log}. 
Redistribution matrix files (rmf) and ancillary response files (arf) were generated with the SAS tasks {\sc rmfgen} and {\sc arfgen}, and were binned in order to over-sample the instrumental resolution by at
least a factor of four, with no impact on the fit results.
We notice that for the arf calculation, we applied the new option {\sc applyabsfluxcorr=yes} which allows for a correction of the order of 6–8\% between 3 and 12 keV in order to reduce differences in the spectral shape between {\sl XMM-Newton}-pn and {\sl NuSTAR} spectra (F.\ Fürst 2022, XMM-CCF-REL-388, XMM-SOC-CAL-TN-0230)\footnote{https://xmmweb.esac.esa.int/docs/documents/CAL-SRN-0388-1-4.pdf}.
 Finally, the background-corrected pn spectra were binned in order to have a signal-to-noise  ratio greater than four in each spectral channel.

 The UV data from the XMM-Newton Optical-UV Monitor (hereafter OM; \citealt{Mason01}) were processed using the SAS script {\sc omichain}. This script
 takes into account all necessary calibration processes (e.g. flat-fielding) and runs a source detection algorithm before performing aperture photometry (using an extraction radius of 5$\farcs$7) on each detected
source, and combines the source lists from separate exposures
into a single master list in order to compute mean corrected count rates.
In order to take into account the OM calibration uncertainty
of the conversion factor between the count rate and the flux, we
quadratically added a representative systematic error of 1.5\%\footnote{https://xmmweb.esac.esa.int/docs/documents/CAL-SRN-0378-1-1.pdf} to the statistical error of the count rate, as done in \cite{Porquet19} for the SED analysis of the bare AGN Ark\,120.\\

{\sl NuSTAR} \citep{Harrison13} observed Mrk\,110 with its two
co-aligned X-ray telescopes with corresponding Focal Plane Modules A 
(FPMA) and B (FPMB).  The level 1 data products were
processed with the {\sl NuSTAR} Data Analysis Software (NuSTARDAS)
package (v2.1.2, released on March 14 2022). Cleaned event files (level 2 data products) were
produced and calibrated using standard filtering criteria with the
\textsc{nupipeline} task and the calibration files available in the
{\sl NuSTAR} calibration database (CALDB version: 20220829).
Extraction radii for both the source and the background spectra were 60 arcseconds.
The corresponding net exposure time for the observations with the FPMA and FPMB
are reported in Table~\ref{tab:log}. 
 The processed rmf and arf files were provided on a linear grid of 40 eV steps. 
As the FWHM energy resolution of {\sl NuSTAR} is 400\,eV below $\sim$50 keV and increases to 1\,keV at 86\,keV \citep{Harrison13}, 
we re-binned the rmf and arf files in energy and channel space by a factor of 4 to over-sample the instrumental energy resolution
by at least a factor of 2.5.  
The background-corrected {\sl NuSTAR} spectra were finally binned in order to have a signal-to-noise ratio greater than four in each spectral channel.

\subsection{Spectral analysis method}\label{sec:method}

The {\sc xspec} v12.12.1 software package \citep{Arnaud96}
 was used for the spectral analysis.
As found by \cite{Reeves21b}, there is no additional X-ray absorption compared to the Galactic value,
which we fixed to 1.27$\times$10$^{20}$\,cm$^{-2}$ \citep{HI4PI16}.
We applied the X-ray absorption model {\sc tbnew (version 2.3.2)} from
\cite{Wilms00}, setting their interstellar medium (ISM) elemental abundances and using
the cross-sections from \cite{Verner96}. 
We allowed for cross-calibration uncertainties between the two {\sl NuSTAR} spectra and the {\sl XMM-Newton}-pn spectra in the fit by including a cross-normalisation factor  
for the pair of {\sl NuSTAR} FPMA and FPMB spectra, with respect to the pn spectra.  
We used $\chi^{2}$ minimisation throughout, quoting errors with 90\% confidence intervals 
for one interesting parameter ($\Delta\chi^{2}$=2.71). Default values of H$_{\rm 0}$=67.66\,km\,s$^{-1}$\,Mpc$^{-1}$, $\Omega_{\rm m}$=0.3111, and $\Omega_{\Lambda}=0.6889$ were assumed
\citep{Planck20}.

\section{X-ray simultaneous broadband spectral analysis with the {\sc ReXcor} model: probing the physical origin of the soft X-ray excess}\label{sec:rexcor}

 {\sc ReXcor} is a new phenomenological X-ray (0.3--100 keV) spectral fitting model of the disc-corona system in AGN that self-consistently combines the effects of both the emission for a warm corona and ionised relativistic reflection \citep{Xiang22}, which is based on the procedure described in \cite{Ballantyne20a} and \cite{Ballantyne20b}. The accretion energy released in the inner disc is apportioned between the three system components (see Fig.\,1 in \citealt{Xiang22}): the warm corona, the hot corona (assuming a lamppost geometry,  located above the spin axis of the black hole), and the accretion disc.
 This model includes the effects of relativistic light-bending and blurring up to 400\,$R_{\rm g}$ using the {\sc relconv$\_$lp} convolution model \citep{Dauser13}, assuming isotropic limb darkening. Depending on the fraction of energy dissipated in the warm and hot coronae, as well as the warm corona heating fraction and optical depth, various soft X-ray excess shapes are produced (see Figures.~4 and 5 in \citealt{Xiang22}). The assumed metal abundances are from \cite{Morrison83}.\\

 The eight publicly available table grid models are provided for two black hole spin values ($a$=0.90 and $a$=0.99), two hot-corona height values ($h$=5\,$R_{\rm g}$ and $h$=20\,$R_{\rm g}$), and two Eddington-ratio values ($\dot{m}=$0.1 and $\dot{m}$=0.01). These grids were computed for a disc inclination angle of 30 degrees. Therefore, in the present X-ray broadband analysis, we are not able to explore the impacts on the fit results using differing values. However, here our aim is primarily to estimate the contribution of a warm corona (compared to relativistic reflection) to the soft X-ray excess.\\

 The free model parameters of the {\sc reXcor} grids are:\begin{itemize}
  \item the hot-corona heating fraction with 0.02$\leq$$f_{\rm X}$$\leq$0.2;
  \item the photon index of irradiating power law from the hot corona with 1.7$\leq$$\Gamma$$\leq$2.2;
  \item the warm-corona heating fraction with 0.0$\leq$ $h_{\rm f}$$\leq$0.8: $h_{\rm f}$=0 means that the soft X-ray excess is exclusively due to relativistic reflection;
  \item the warm-corona Thomson depth with 10$\leq$$\tau$$\leq$30.
 \end{itemize}

 The detailed fitting of the baseline model and the best-fit results for the four model grids are reported in Appendix~\ref{sec:rexcorapp}. The values of the warm-corona heating fraction are high ($h_{\rm f}$$\sim$50--70\%), indicating that the soft X-ray excess is mainly produced by a warm corona. This supports our previous results, obtained using a simplified Comptonisation model ({\sc comptt}) plus relativistic reflection. We note that lower BH spin and/or larger coronal height values (which are not provided in the current model grids) would lead to a weaker reflection fraction \citep[e.g.,][]{Dauser14}. This would then require a stronger contribution from the warm corona in order to reproduce the soft X-ray excess for Mrk\,110.
   Therefore, the two BH spin values (and the two hot-corona-height values) investigated here could be considered as a conservative estimate of the warm corona contribution for Mrk\,110.
 The warm-corona optical depth is rather high with $\tau_{\rm warm}$$\sim$13--28. For both epochs, only a few percent of the accretion energy ($f_{\rm X}$$\sim$3--7\%) is dissipated in the lamppost hot corona.

 The 0.3--10\,keV flux ratios of the {\sc reXcor} and {\sc zcutoffpl} components strongly diverge from unity, with flux({\sc rexcor})/flux({\sc zcutoffpl})$\sim$0.23. As discussed by \cite{Xiang22}, this could be due to effects not included in their models, such as, for example, a truncated accretion disc.
 This would lead to reduced emission from the warm corona and a reduction in the relativistic reflection fraction.
  Since the {\sc reXcor} model grids set the inner radius of the accretion disc to the inner stable circular orbit value (ISCO), we cannot perform the study assuming a truncated accretion disc.
  However, this truncated accretion disc scenario is investigated in the next section when modelling the spectral energy distribution (UV to hard X-rays) for both epochs with the {\sc relagn} model.

\section{SED analysis from UV to hard X-rays of Mrk\,110 with the {\sc relagn} model: disc-corona system properties from its energetics}\label{sec:SED}

We now investigate the SED of Mrk\,110 from UV to hard X-rays to determine the physical properties of its disc-corona system.
For this purpose, we use the new {\sc relagn} model, which is based on the {\sc agnsed} code of \cite{Kubota18}, but which incorporates general relativistic ray-tracing \citep{Hagen23b}. The model consists of an inner optically-thin hot corona ($R_{\rm ISCO}$$\leq$$R$$\leq$\,$R_{\rm hot}$), a warm Comptonised disc ($R_{\rm hot}$\,$\leq$$R$$\leq$\,$R_{\rm warm}$) and an outer standard disc ($R_{\rm warm}$$\leq$$R$$\leq$$R_{\rm out}$). An illustration of this disc-corona geometry is displayed in \cite{Kubota18} (their fig.~2). It is worth noting that for the {\sc relagn} model the disc is truncated below $R_{\rm hot}$; whereas for the {\sc reXcor} model ($\S$\ref{sec:rexcor}) the inner accretion disc radius is set to ISCO with a lamppost geometry for the hot corona.

\begin{figure*}[t!]
\begin{tabular}{cc}
\includegraphics[width=0.95\columnwidth,angle=0]{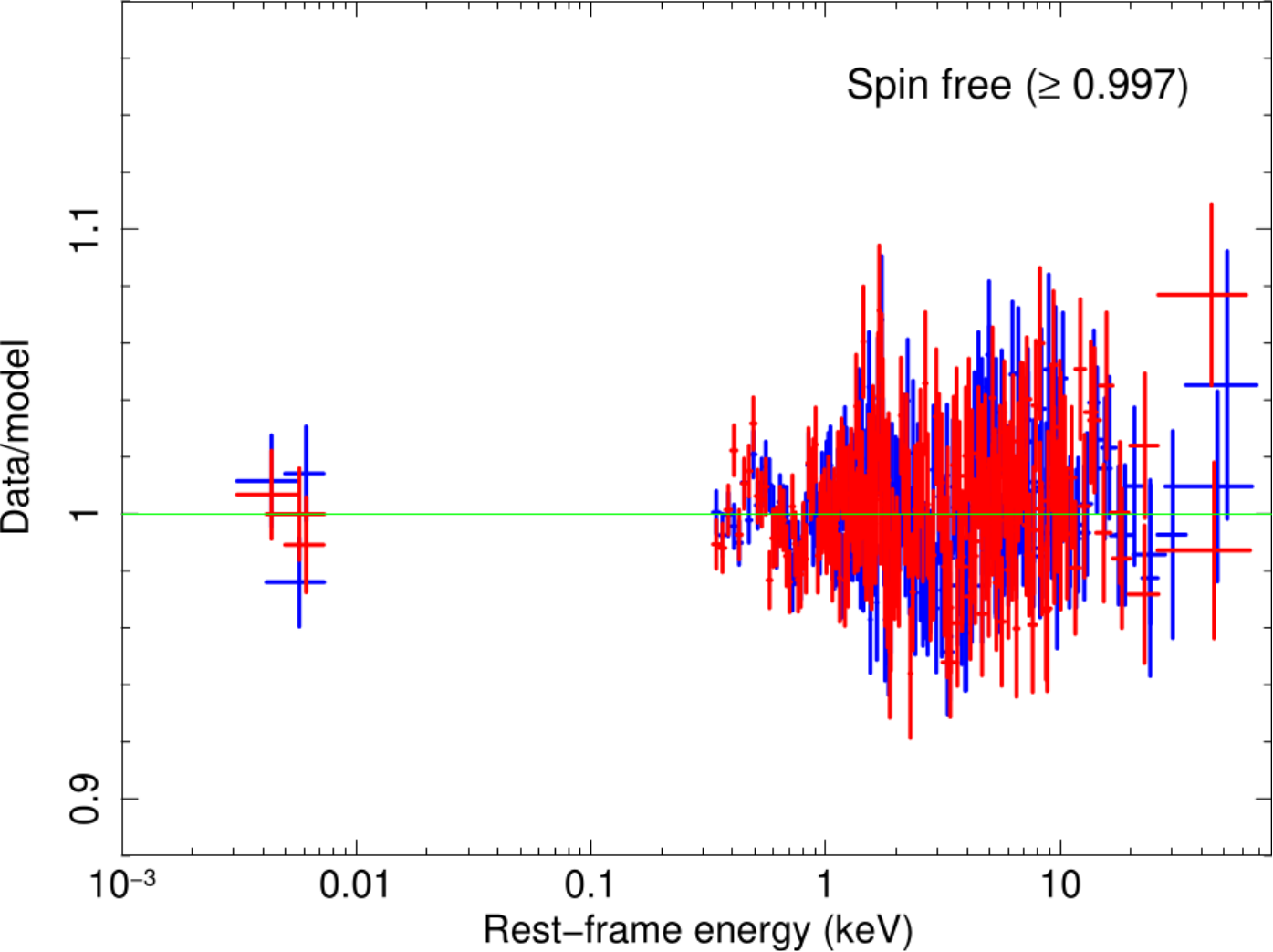} &
\includegraphics[width=0.95\columnwidth,angle=0]{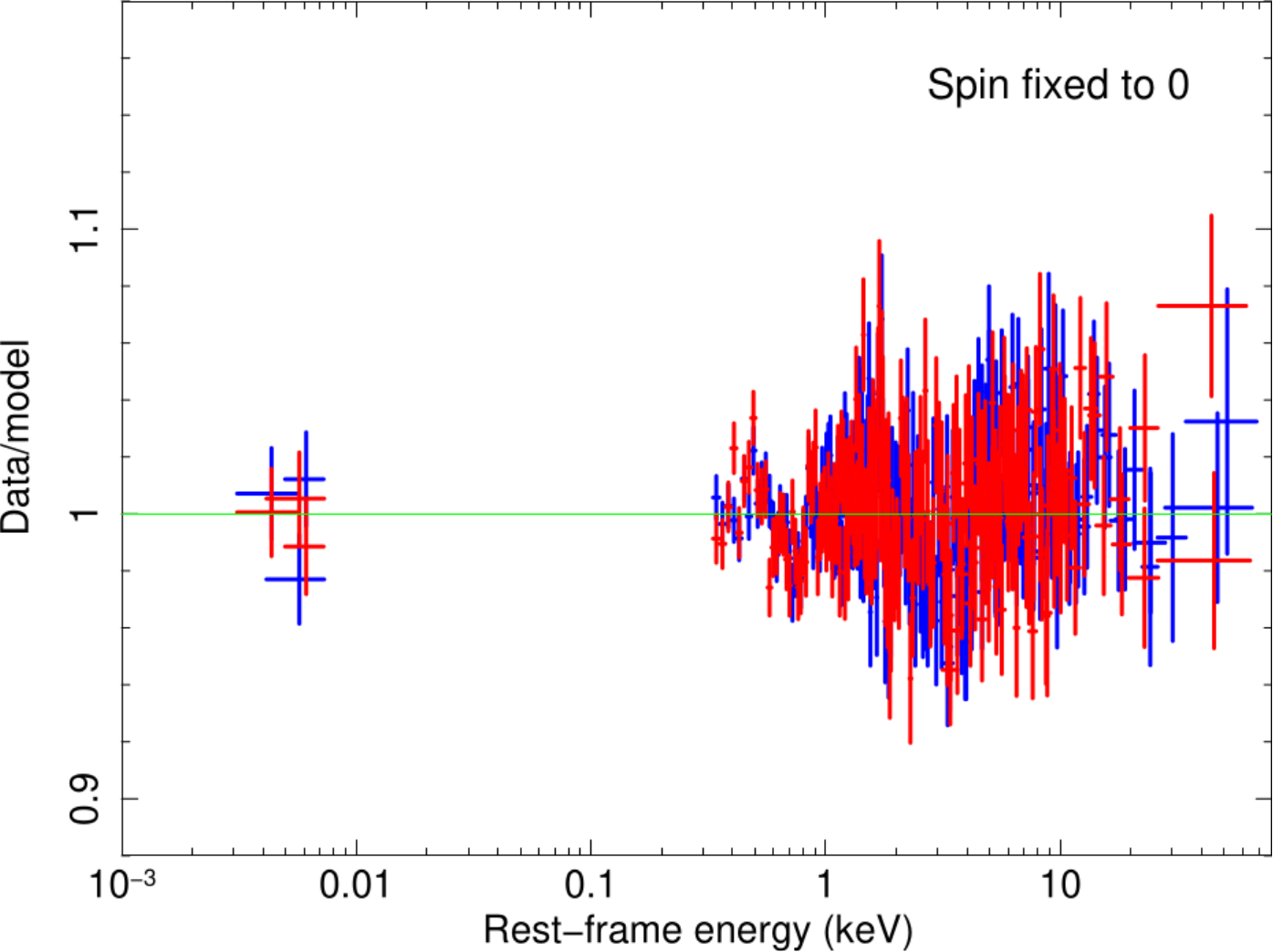} \\
\includegraphics[width=0.95\columnwidth,angle=0]{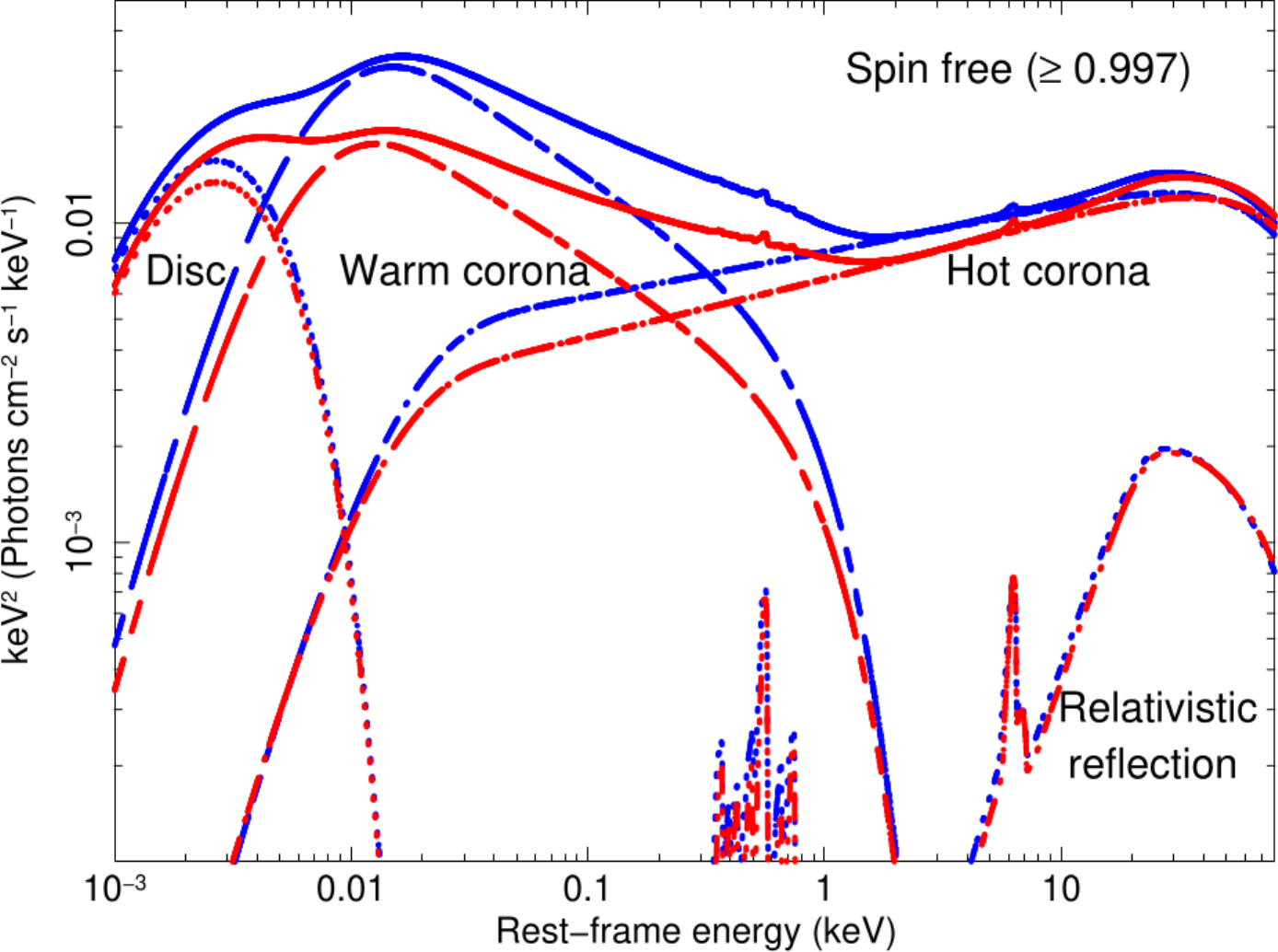} &
\includegraphics[width=0.95\columnwidth,angle=0]{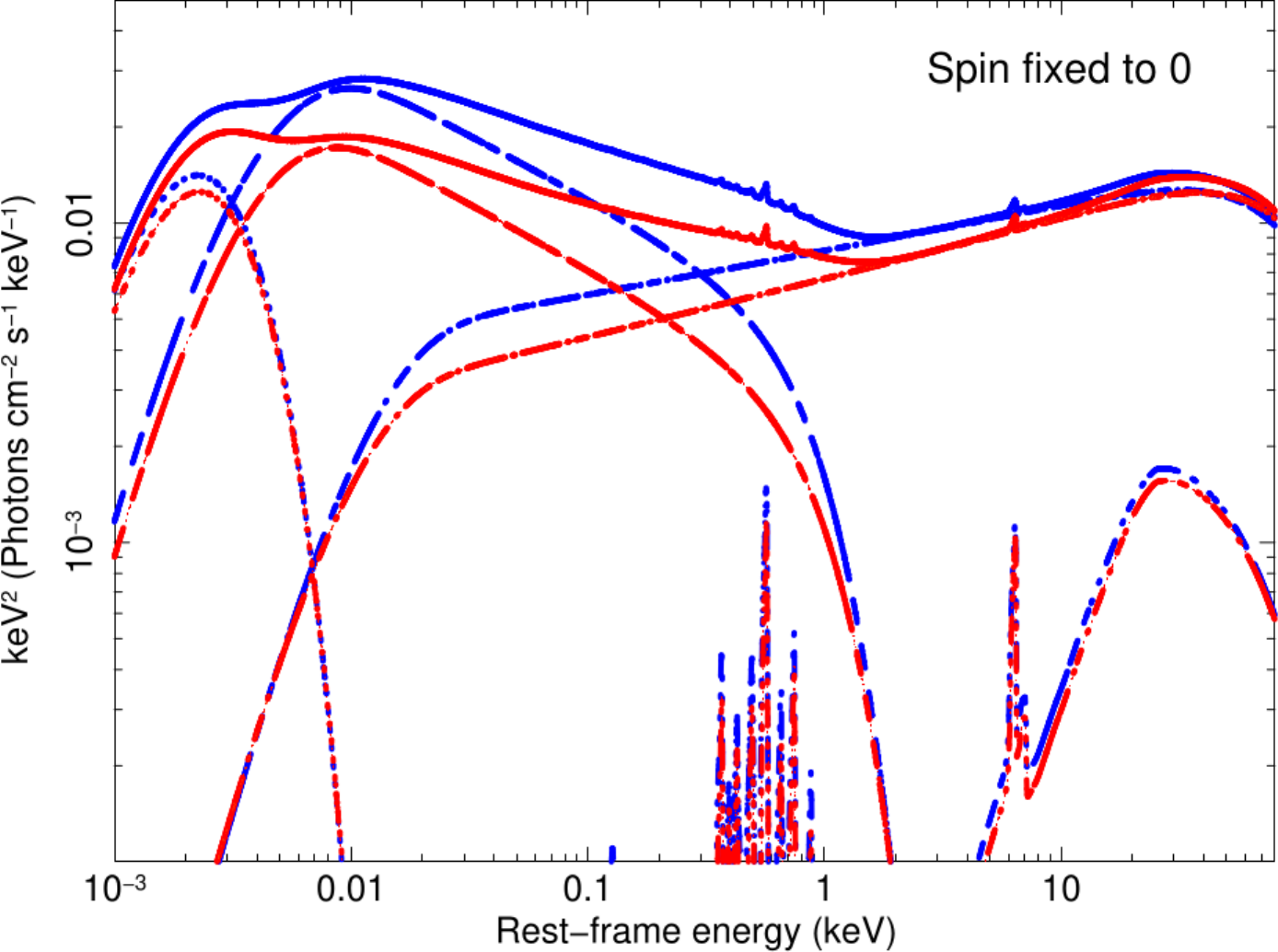} \\
\end{tabular}
\caption{SED fit from UV to hard X-rays of Mrk\,110 using the {\sc relagn+relxillcp} baseline model for the 2019 (blue) 2020 (red) simultaneous {\sl XMM-Newton} and {\sl NuSTAR} spectra. Left panels: the black-hole spin is allowed to vary, and is found to be greater or equal to 0.997. The values of the best-fit parameters are reported in Table~\ref{tab:SED}. Right panels: the black-hole spin value is fixed to zero.  The best-fit parameter values are reported in Table~\ref{tab:SEDspin}. Top panels: Data-to-model ratio. Bottom panels: intrinsic SED corrected for Galactic reddening  and absorption (solid curves), with the main individual emission components of the baseline model: outer disc (dotted curves), warm, optically-thick Comptonisation (dashed curves; warm corona), hot, optically-thin Comptonisation (dotted-dashed curves; hot corona), and relativistic reflection (three-dotted-dashed curves).}
\label{fig:SED}
\end{figure*}

 \begin{table}[t!]
\caption{Fits of the simultaneous 2019 and 2020 SED (UV to hard X-rays) of Mrk\,110
 with the {\sc relagn+relxillcp} baseline model.}
\centering
\begin{tabular}{@{}l c c}
\hline\hline
parameter & \multicolumn{1}{c}{2019 Nov}  & \multicolumn{1}{c}{2020 April}\\
\hline
$a$ & \multicolumn{2}{c}{$\geq$0.997}  \\
log\,$\dot{m}$ & $-$1.03$^{+0.01}_{-0.03}$ & $-$1.14$^{+0.01}_{-0.03}$ \\
$kT_{\rm hot}$ (keV)   & \multicolumn{2}{c}{58$^{+25}_{-8}$} \\
$\Gamma_{\rm hot}$ & 1.86$\pm$0.01 & 1.82$\pm$0.01  \\
$R_{\rm hot}$ (R$_{\rm g}$) & 16$^{+1}_{-4}$ & 20$\pm$1 \\
$kT_{\rm warm}$ (keV) & 0.23$\pm$0.01 & 0.24$\pm$0.01 \\
$\Gamma_{\rm warm}$ & 2.48$^{+0.02}_{-0.03}$  & 2.46$^{+0.03}_{-0.04}$ \\
$R_{\rm warm}$  (R$_{\rm g}$)& 88$^{+4}_{-3}$ & 79$^{+3}_{-8}$ \\
log\,$\xi$  & \multicolumn{2}{c}{1.0$\pm$0.2} \\
norm(relxillcp) & 2.1$^{+0.6}_{-0.5}$$\times$10$^{-5}$ & 1.9$^{+0.4}_{-0.5}$$\times$10$^{-5}$ \\
F(0.3--79\,keV)$^{(a)}$ &   11.8$\times$10$^{-11}$ &   9.7$\times$10$^{-11}$ \\
L(0.3--79\,keV)$^{(b)}$ &   3.1$\times$10$^{44}$ &  2.7$\times$10$^{44}$ \\
$\chi^{2}$ (d.o.f.; $\chi^{2}_{\rm red}$) & \multicolumn{2}{c}{1725.0 (1614; 1.07)}  \\
\hline    \hline
\end{tabular}
\label{tab:SED}
\flushleft
\small{\textit{Notes}. $^{(a)}$ Observed fluxes (erg\,cm$^{-2}$\,s$^{-1}$).
$^{(b)}$ Intrinsic luminosities (erg\,s$^{-1}$).}
\end{table}

The model parameters of {\sc relagn} are identical to those of {\sc agnsed}, except an additional parameter that allows for a colour-temperature correction\footnote{In {\sc agnsed}, the colour-temperature correction is hardwired at unity.} to the standard outer disc ($f_{\rm col}$). For a detailed description of the {\sc relagn} model we refer to \cite{Hagen23b}. We note that in the case where relativistic effects are not taken into account ({\sc agnsed}), the spin and accretion rate would be significantly underestimated for Mrk\,110 (see Table~\ref{tab:SEDfits}), as also pointed out by \cite{Hagen23a} for \object{Fairall\,9}.

We only use data from the three shortest-wavelength UV filters with the OM (UVW2, UVM2, UVW1; effective wavelengths: 2\,120, 2\,310, 2\,910\,\AA, respectively) since the contaminations by the host galaxy and the close foreground star located at 5$\farcs$1-NE are negligible in these bands (Lobban et al., 2023, MNRAS to be subm.; hereafter L23). From monitoring over several months of Mrk\,110 with {\sl Neil Gehrels Swift} observatory the time delay between the X-rays and the UV emission is consistent with zero lag with an upper limit of $\sim$1 day (\citealt{Vincentelli21}, L23). Moreover, the UV and X-rays fluxes are not observed to vary significantly above their statistical errors during the time-elapsed durations of the 2019 and 2020 {\sl XMM-Newton} ($\sim$8\,hours) and {\sl NuSTAR} ($\sim$1\,day) observations (L23). Therefore, the UV and X-ray emission can be considered to be  effectively simultaneous on this timescale, which is an important assumption for the SED modelling.

The Galactic reddening of Mrk\,110 is very low: $E$($B-V$) = 0.01 \citep{Schlafly11}.
For the colour-correction of the outer disc, we apply the relation in \cite{Done12}, by setting $f_{\rm col}$$<$0 in the model.
For each epoch, we tied the upper limit of the scale heights  of the hot corona component to the $R_{\rm hot}$ value ($h_{\rm max}$ in $R_{\rm g}$), but we also checked that fixing it, for example, to 10\,$R_{\rm g}$ does not impact our results.
The distance of Mrk\,110 is fixed to 155.7\,Mpc \citep{Wright06,Planck20}, its black hole mass is fixed to 1.4$\times$10$^{8}$\,M$_{\odot}$ \citep{Kollatschny03,Porquet21}, and its accretion disc inclination is fixed to 10 degrees \citep{Reeves21b}.

Since a component of mild relativistic reflection is present \citep{Reeves21b,Porquet21} - but not included in the {\sc relagn} model - we also add the {\sc relxillcp} model, which uses an underlying Comptonisation continuum and a broken emissivity for the hot corona \citep[][version 2.2]{Dauser14}. The emissivity indices are both fixed to the canonical values of three. Since the contribution of the relativistic reflection to the spectra is much weaker than that of {\sc relagn} (see Fig.~\ref{fig:SED}, bottom panels), its real shape has no significant impact on the results. Indeed, similar results are found for the relativistic reflection component by assuming a lamppost geometry for the hot corona using {\sc relxilllpcp} (see Appendix~\ref{app:diffsed}, Table~\ref{tab:SEDfits}).
The inner radius of the relativistic-reflection component is set to the $R_{\rm hot}$ radius since the disc is truncated below this value.\\

Our baseline model is: {\sc tbnew$\times$redden(relagn + relxillcp + zgaussian)}. As shown in Fig.~\ref{fig:SED} (top left panel), a very good fit from UV to hard X-rays is obtained with the physical-parameters values reported in Table~\ref{tab:SED}. Mrk\,110 has a moderate accretion rate and its value (in log scale) decreases from $-$1.03$^{+0.01}_{-0.03}$ to $-$1.14$^{+0.01}_{-0.03}$ (at 3.5$\sigma$ confidence level), with the X-ray flux decreasing by only a factor of 1.13.  The properties of the warm corona ($kT_{\rm warm}$ and $\Gamma_{\rm warm}$) are very similar between both epochs. There is a hint of a decrease of the mean value of $R_{\rm warm}$ when the source flux increases, though the values are compatible within their error bars calculated at 90\% confidence level. For the hot corona, there is a slight spectral hardening and a decrease of the hot corona radius when the source flux increases (at 2.8$\sigma$ confidence level for both). This latter trend is similar to that found for the bare AGN Ark\,120, which also accretes at a moderate Eddington rate \citep{Porquet18,Porquet19}. The outer radius of the hot corona ($R_{\rm hot}$) is consistent with that found from the variability of the mildly-relativistic soft X-ray \ion{O}{vii} lines \citep{Reeves21b}.

\begin{table*}[t!]
\caption{Simultaneous SED fits of the 2019 and 2020 simultaneous {\sl XMM-Newton}-pn and {\sl NuSTAR} spectra of Mrk\,110.
The model is: {\sc tbnew(Gal)$\times$redden(Gal)(relagn+relxillcp+zgaussian)}, fixing the black hole spin value.
 `(t)' means that the value has been tied between both epochs.
 `(f)' means that the values has been fixed.
}
\centering
\begin{tabular}{@{}l c c c c c}
\hline\hline
parameter & $a$=0 & $a$=0.5 & $a$=0.7 &$a$=0.90 & $a$=0.998 \\
\hline
\multicolumn{6}{c}{2019}\\
\hline
log\,$\dot{m}$              &  $-$1.37$\pm$0.01 & $-$1.34$\pm$0.01  & $-$1.31$\pm$0.01 & $-$1.24$\pm$0.01 & $-$1.03$\pm$0.01\\
$kT_{\rm hot}$ (keV)   & 42$^{+17}_{-9}$  & 43$^{+17}_{-8}$  & 44$^{+18}_{-8}$ &  46$^{+22}_{-9}$  & 57$^{+29}_{-13}$ \\
$\Gamma_{\rm hot}$ & 1.86$\pm$0.01 & 1.86$\pm$0.01  & 1.86$\pm$0.01 & 1.86$\pm$0.01 & 1.86$\pm$0.01 \\
$R_{\rm hot}$ (R$_{\rm g}$) & 43$\pm$1 & 32$\pm$1 & 27$\pm$1 & 21$\pm$1 & 16$\pm$1  \\
$kT_{\rm warm}$ & 0.20$\pm$0.01 & 0.21$\pm$0.01 & 0.21$\pm$0.01 & 0.22$\pm$0.01 & 0.23$\pm$0.01 \\
$\Gamma_{\rm warm}$ & 2.38$\pm$0.01  & 2.41$\pm$0.01 & 2.43$\pm$0.02 & 2.45$\pm$0.02 & 2.48$\pm$0.03\\
$R_{\rm warm}$  (R$_{\rm g}$)& 199$^{+19}_{-15}$ & 159$^{+15}_{-12}$ & 139$^{+13}_{-11}$ & 112$^{+11}_{-10}$ & 87$\pm$9 \\
log\,$\xi$ & 1.0 (f) & 1.0 (f) & 1.0 (f) & 1.0 (f) & 1.0 (f)\\
norm(relxill) & 1.9$\pm$0.5$\times$10$^{-5}$ & 1.9$\pm$0.5$\times$10$^{-5}$ & 2.0$\pm$0.5$\times$10$^{-5}$ & 2.1$\pm$0.5$\times$10$^{-5}$ &  2.2$\pm$0.6$\times$10$^{-5}$ \\
\hline
 \multicolumn{6}{c}{2020}\\
\hline
log\,$\dot{m}$  & $-$1.46$\pm$0.01 & $-$1.43$\pm$0.01  & $-$1.40$\pm$0.01 & $-$1.34$\pm$0.01 & $-$1.14$\pm$0.01 \\
$kT_{\rm hot}$ (keV)   & (t) & (t) & (t) & (t) & (t) \\
$\Gamma_{\rm hot}$ & 1.82$\pm$0.01 & 1.82$\pm$0.01 & 1.82$\pm$0.01 & 1.82$\pm$0.01 & 1.82$\pm$0.01  \\
$R_{\rm hot}$ (R$_{\rm g}$) & 51$^{+2}_{-1}$ & 39$\pm$1  & 33$\pm$1 & 26$\pm$1 & 20$\pm$1\\
$kT_{\rm warm}$ & 0.22$\pm$0.01 & 0.22$\pm$0.01 & 0.23$\pm$0.01 & 0.23$\pm$0.01 & 0.24$\pm$0.01   \\
$\Gamma_{\rm warm}$  & 2.39$\pm$0.01 & 2.41$\pm$0.02 & 2.42$\pm$0.02 & 2.44$^{+0.02}_{-0.03}$ & 2.46$\pm$0.03  \\
$R_{\rm warm}$  (R$_{\rm g}$) & 179$^{+15}_{-12}$ & 144$^{+11}_{-10}$ & 126$^{+10}_{-9}$ & 101$^{+9}_{-8}$& 79$^{+8}_{-7}$\\
log\,$\xi$ & (t) & (t) & (t) & (t) & (t) \\
norm(relxill) & 1.5$\pm$0.5$\times$10$^{-5}$ & 1.6$\pm$0.5$\times$10$^{-5}$ & 1.7$\pm$0.5$\times$10$^{-5}$ & 1.8$\pm$0.5$\times$10$^{-5}$ & 1.9$\pm$0.5$\times$10$^{-5}$\\
\hline
$\chi^{2}$/d.o.f. & 1750.1/1612 & 1741.6/1612 & 1736.7/1612 & 1730.0/1612 & 1724.9/1612  \\
$\chi^{2}_{\rm red}$ & 1.09 & 1.09 & 1.08 & 1.07 & 1.07 \\
\hline    \hline
\end{tabular}
\label{tab:SEDspin}
\end{table*}

The best-fit spin value is found to be extreme with a lower limit of 0.997. However, we check whether different values of the spin can be really excluded by fixing it to values of 0, 0.5, 0.7, and 0.9. We fix the ionisation parameter ($\xi$; erg\,cm\,s$^{-1}$) of the accretion disc (in log scale) to unity in order to allow for the presence of the \ion{O}{vii} disc lines \citep{Reeves21b}. As shown in Table~\ref{tab:SEDspin}, the $\chi^{2}$ value (d.o.f.=1612) increases with the decrease of the spin value, by up to about $\Delta\chi^{2}$=+25 for a zero-spin value. However, the fit is still satisfactory (Fig.~\ref{fig:SED}: top right panel), indicating that even if the best-fit is found for a maximally-spinning black hole, a non-spinning black hole cannot be ruled out. For a black-hole spin value of 0, the accretion rate would decrease down to $\sim$3--4\%, while the radii of the hot and warm coronae would increase by about a factor of two up to $\sim$50\,$R_{\rm g}$ and $\sim$200\,$R_{\rm g}$, respectively. Comparing the SED shape between a non-spinning black hole (Fig.~\ref{fig:SED}: bottom-right panel) and a maximally-spinning black hole (Fig.~\ref{fig:SED}: bottom-left panel), the emission of the warm corona is slightly weaker and is shifted to lower energies as expected from the increase of the outer hot corona radius, or in other words, to the increase of the inner warm-corona radius.
 This is due to the energy balance used in {\sc relagn}. As the BH spin decreases, $R_{\rm hot}$ must increase in order to compensate for the smaller emitting area (due to in increase of the ISCO) and lower accretion efficiency. This in turn leads to a necessary increase in $R_{\rm warm}$, again to compensate for a smaller emitting area and lower efficiency. See \cite{Hagen23b} for a detailed discussion.

\section{Discussion}\label{sec:discussion}

In previous work reported by \cite{Porquet21}, we have demonstrated that the first ever simultaneous {\sl XMM-Newton} and {\sl NuSTAR} spectrum of the bare AGN Mrk\,110 -- obtained in 2019 (November 16--17) and 2020 (April 5--6) -- cannot be reproduced by purely relativistic reflection. From a simplified model combining soft and hard Comptonisation (using the {\sc comptt} model) and mild relativistic reflection, the broadband X-ray continuum is very well reproduced. Therefore, in this work our first aim is to confirm with a more sophisticated model if the origin of the soft X-ray excess observed in Mrk\,110 can physically originate from a warm corona. To test this, the {\sc reXcor} model is applied to the X-ray broadband spectra above 0.3\,keV. This model self-consistently combines the effects of a warm corona with the X-ray relativistic reflection and allows for a physical estimate of the warm-corona heating fraction \citep{Xiang22}.
Then, we perform an in-depth SED analysis (from UV to hard X-rays) using the {\sc relagn} model \citep{Hagen23b} -- adding a relativistic reflection component -- to infer the accretion rate and the physical properties of the disc-corona system. The main spectral analysis results are summarised and discussed below.

\subsection{The warm corona as the main origin of the soft X-ray excess}

The {\sc reXcor} model was applied to the 2019 and 2020 simultaneous X-ray broadband spectra of Mrk\,110, noting the limitations of the range of paramer space available within the model grids (see details in $\S$\ref{sec:rexcor}). For both epochs, the warm-corona heating fraction ($h_{\rm f}$) is large: $\sim$50--70\%, corroborating that the soft X-ray excess of Mrk\,110 mainly originates from a warm corona rather than relativistic reflection  \citep{Porquet21}.  The high values of its optical depth ($\tau_{\rm warm}$$\sim$13--28) provide a smooth soft X-ray excess and are consistent with very recent modelling using a disc-corona structure, taking into account both magnetic and radiation pressure for an accretion rate of about 0.1 \citep{Gronkiewicz23}.

Only a few percent of the accretion energy ($f_{\rm X}$$\sim$3--6\%) is dissipated in the lamppost hot corona for Mrk\,110, as also found by \cite{Xiang22} for \object{HE 1143-1820} and \object{NGC\,4593}. Such a low value for $f_{\rm X}$ allows for a significant relative strength of the soft excess. Indeed, as discussed by \cite{Xiang22}, at such a value, the irradiated gas is less ionised, leading to significant absorption between about 1--4\,keV \citep[see fig.~2 in][]{Ross99},  and then to a stronger observed contrast between the soft X-ray excess and the higher-energy emission.
 Our best-fit result would favour a spin value of 0.99 for Mrk\,110, compared to the other available grids calculated with a black hole spin of 0.90. But, the model grids are only calculated for these two spin values (0.90 and 0.99) preventing us from investigating the results for other spin values for Mrk\,110.

The low 0.3--10\,keV fluxes of the {\sc reXcor} model - which only includes the emission and reflection component - compared to  the primary continuum could suggest a truncated accretion disc, leading to weakened component of reflection emission.
 This truncated accretion disc scenario is further investigated when modelling the SED (UV to hard X-rays) of Mrk\,110 for both epochs with the {\sc relagn} model.

\subsection{The properties of the disc-corona system of Mrk\,110 inferred from its SED analysis}

We analysed the 2019 and 2020 spectral energy distribution from UV to hard X-rays ($\S$\ref{sec:SED}), adopting the new model {\sc relagn} based on the {\sc agnsed} model \citep{Kubota18}, but including relativistic ray-tracing \citep{Hagen23b}. The mild relativistic reflection is taken into account by adding the {\sc relxillcp} model \citep{Dauser14}.

This model reproduces the SED of Mrk110 at both epochs very well (Fig.~\ref{fig:SED}).
 The best-fit is found for an extreme black hole spin with a lower limit of 0.997. However, other spin values cannot be excluded. Indeed, for the case of a zero black hole spin the fit is still statistically satisfactory. The outer radii of the hot and warm coronae (depending on the black hole spin value) are located at a few 10s and $\sim$100 $R_{\rm g}$, respectively.\\

The measured Eddington rate range of Mrk\,110 in 2019 and 2020 is about only 3--9\% (depending on its flux level and the black hole spin), which is similar to that of the two BLS1 bare AGN Ark\,120 \citep[$\sim$3\% in 2013 and $\sim$7\% in 2014;][]{Porquet19} and Fairall\,9 \citep[$\sim$10\% in 2014;][]{Hagen23a}.
 Though Mrk\,110 exhibits a 10\,keV X-ray flux (at both epochs) similar to that observed for Ark\,120 and Fairall\,9, its UV peak emission is weaker by a factor of about 2--5 and 3--5 compared to these two latter objects, respectively \citep{Porquet19,Hagen23a}.
In order to determine the physical origin(s) of such different SED shapes, a thorough comparison of their spectral and timing disc-corona properties and black hole spin is necessary. This is beyond the scope of this article and will be presented in a forthcoming work. \\
We found a trend of a hard X-ray spectral hardening and a decrease in the radius of the hot corona when the source flux increases, as found for the bare AGN Ark\,120 \citep{Porquet19}. The photon index of the hot corona of Mrk\,110 displays a slight hardening between 2019 (highest X-ray flux) and 2020 (lowest X-ray flux). This is consistent with the `softer-when-brighter' behaviour commonly observed in type-I AGN with accretion rate above about 1\% \cite[e.g.,][]{Markowitz03,Porquet04a,Younes11,Soldi14,Connolly16,Ursini16,Gliozzi17,Weng20}. Importantly, the photon index values are found to be about 1.8--1.9, consistent with previous X-ray observations of Mrk\,110 and, more generally, with BLS1s \citep[e.g.,][]{Porquet04a,Zhou10,Waddell20,Gliozzi20}.\\

As with any modelling, there are some possible caveats. The disc-corona geometry in AGN is not established and could be different to that assumed by the {\sc relagn} model.
Additionally, a disc-like geometry is supposed when incorporating the relativistic ray-tracing, while the hot-corona geometry could have a much larger scale height, as found, for example, for the zero black-hole spin value where the maximum height of the corona is the highest ($\sim$40\,$R_{\rm g}$). Therefore, we perform the SED spectral fit for a black-hole spin of zero, but fixing the maximum height of the corona height to 10\,$R_{\rm g}$, rather than tying it to $R_{\rm hot}$ - however, no noticeable impact on the {\sc relagn} parameter values is found.

In the SED analysis, the UV emission is supposed to come exclusively from the disc region - however, other contributions may be present. For example, in many AGN, an excess in the U-band (around the Balmer jump at 3\,465\AA) continuum lags by about a factor of two - compared to an extrapolation of the trend through the rest of the UV/optical regime - is observed  \citep{Cackett18,Edelson19,Cackett20,Homayouni22}. This excess can be explained by the significant `diffuse continuum' from the BLR itself or from the wind inwards of the BLR  \citep{Korista01,Korista19,Lawther18,Dehghanian19b,Mahmoud23,Hagen23b}. Such a U-band continuum-lag excess is also observed in Mrk\,110 when observed in relatively high-flux states \citep{Vincentelli22}, but not at lower-flux states \citep{Vincentelli21}, which are similar to the present 2019 and 2020 {\sl XMM-Newton} and {\sl NuSTAR} observations, respectively. Since the UVW1 filter band (which overlaps with the U band)  includes the Balmer jump, the UV emission from the outer disc could be overestimated. Removing it from the SED fits only leads to a slight increase of the outer warm-corona radius (compatible with previous values within the error bars), and has a negligible impact on the other parameters.

\subsection{The long-term variability of Mrk\,110 combined with its low accretion rate: indication for a (moderate) changing-state AGN ?}

From the Sloan Digital Sky Survey (SDSS) and Pan-STARRS1 \citep{MacLeod16}, about 1\% of their AGN sample display variability amplitude in the $g$ band of at least one magnitude on time-scales shorter than 15 years, and up to 30--50\% on longer time-scales \citep[see also][]{MacLeod19,Lopez-Navas22,Hon22,Temple23}. These types of AGN, which could be a non-negligible fraction of the overall AGN population, have been named  `changing-look' (CL) AGN. They can experience rapid apparent changes in states, from a type 1 AGN with strong, broad emission lines to a type 2 AGN with only narrow emission lines (no more BLR) - or, conversely. This phenomenon can be explained, for example, by  a significant change in the accretion rate (in this case, these AGN are also called `changing-state' AGN), or in some cases simply by transient obscuration of the BLR. Changing-state behaviour can occur on timescales of only a few years, typically much faster than the viscous timescale from standard accretion theory. Changing-state AGN exhibit lower Eddington ratios relative to the less variable AGN population \citep{MacLeod16,Green22}.\\

Interestingly, Mrk\,110 is known to be a strongly variable source.
In particular, its optical lines and continuum are highly variable \cite[e.g,][]{Bischoff99,Kollatschny01,Veron-cetty07,Vincentelli21,Vincentelli22,Homan23}.
As found from the very long-term ($\sim$30 years; 1987--2019) optical behaviour of Mrk\,110 \citep{Homan23},
the \ion{He}{ii}\,$\lambda$4686 emission line - used as a proxy of the unobservable FUV continuum - displays dramatic variability of a factor of forty and is much higher than the optical continuum.
Also shown in figure 2 from \cite{Homan23}, a very significant drop of the H$\beta$ line flux occured within a very short duration  of $\sim$3--4 years with the lowest state observed in December 2001.
Variable obscuration by intervening dust as the origin of the variability has been ruled out \citep{Homan23}, meaning that the variability of Mrk\,110 is intrinsic to its disc-corona system.
 Moreover, the low Eddington ratio measured for Mrk\,110 is also consistent with what is observed for changing-state AGN, and could explain its relatively steep Balmer decrement of H$\alpha$/H$\beta$ \citep[$\sim$4; ][]{Jaffarian20}.
Indeed, as shown in the recent work by \cite{Wu23} -- based on a photoionisation modelling using the {\it Cloudy} code \citep{Ferland17} taking into account the SED distribution change shapes at different accretion rates -- there is a strong negative correlation between H$\alpha$/H$\beta$ and Eddington ratio is found (see also, \citealt{LaMura07,LuK19}).

During the lowest optical state in December 2001 with SDSS, the broad component of the \ion{He}{ii} line vanished, while
the broad components of H$\alpha$, H$\beta$ and \ion{He}{i}\,$\lambda$5876\AA - though much weaker and narrower - are still present \citep{Homan23}. Since some part of the BLR is still detected, Mrk\,110 did not switch to a Seyfert type 2, but during its lowest optical state could be classified as an intermediate Seyfert type.
Unfortunately, no X-ray data are available during this lowest optical state. The closest-in-time X-ray observation of Mrk\,110 was obtained with {\sl BeppoSAX} in April 2001 where the source was observed in a lower 2--10\,keV flux state ($\sim$ a factor of three) compared to two preceeding {\sl BeppoSAX} observations made in May and November 2000 \citep{Deluit03,Dadina07}, which have similar X-ray fluxes to the present 2019 and 2020 {\sl XMM-Newton} observations. Since the April 2001 X-ray observation of Mrk\,110 occured a few months before its optical lowest flux state, no contemporaneous information within the lag timescale of H$\beta$ \citep[$\sim$25--30\,days;][]{Peterson04} is available to establish if the source was also in a very low X-ray flux state. Contemporaneous/simultaneous multi-wavelength data are crucial, since as shown  from the long-term X-ray light curves (2000-2012) obtained with {\sl RXTE}, Mrk\,110 exhibits significant X-ray flux variations with a flux amplitude of up to about 5 over month-timescales.

Therefore, combining its multi-wavelength spectral and timing characteristics, and the low Eddington ratio inferred from this work, Mrk\,110 could be classified as a (moderate) changing-state AGN. To confirm this scenario, further simultanous/contemporaneous optical/UV/X-rays spectral and timing monitoring of Mrk\,110 at very different flux levels are necessary, especially during its lowest state.

\section{Conclusion}\label{sec:conclusion}

Very high S/N broadband data of bare AGN are the key to probing the disc-corona system and, in particular, to probing hybrid models combining both soft-hard Comptonisation (warm-hot coronae) and relativistic reflection emission.
Here, we applied two brand new codes allowing us to physically take into account the presence of a warm corona, {\sc reXcor} and {\sc relagn}, for the X-ray-bright bare AGN Mrk\,110.

Its simultaneous broadband {\sl XMM-Newton} and {\sl NuSTAR} X-ray spectra in 2019 and 2020 are satisfactorily reproduced by the {\sc reXcor} model. The high values of the warm-corona heating confirm that the soft X-ray excess of Mrk\,110 - for both epochs - originates mainly from a warm corona rather than relativistic reflection \citep{Porquet21}.

 Using the {\sc relagn} model, its UV to hard X-ray SEDs are very well reproduced by the warm- and hot-corona components plus mild relativistic reflection. From the best-fit model with a maximally-rotating SMBH (though other spin values cannot be definitively excluded on a simple statistical basis), the radius of the hot corona is a few 10s\,$R_{\rm g}$, while the warm corona then extends up to $\sim$100\,$R_{\rm g}$.
 For both epochs, the relative strength of the UV compared to X-rays is rather weak, compared to the two other bright bare AGN Ark\,120 and Fairall\,9, which display similar X-ray fluxes. The SED analysis of Mrk\,110 shows that its disc-corona system has a low-to-moderate Eddington ratio of about a few percent. Combined with its long-term optical properties, Mrk\,110 could be classified as a moderate changing-state AGN.

  The success of both the {\sc reXcor} and {\sc relagn} models in fitting the X-ray spectrum and broadband SED of Mrk\,110 is really promising for such disc-corona scenarios including the presence of a warm corona. This provides additional strong evidence that the disc-corona system in AGN can be more complex than the usual scenario assuming purely relativistic reflection from a hot corona with a lamppost geometry onto a standard accretion disc. This reinforces the importance of considering AGN disc-corona systems departing from the standard accretion disc theory  \cite[e.g.,][]{Mitchell23,Temple23b,Hagen23a} and to continue developing and improving self-consistent models incorporating both hot-warm coronae and relativistic reflection.

  The growing number of highly-variable AGN (over time-scales of a few years) which are like Mrk\,110 challenges the standard accretion theory and demonstrates the importance of long-term simultaneous/contemporaneous multi-wavelength monitoring on both single targets and large AGN surveys \citep[e.g.,][]{Green22,Kovacevic22,Kynoch23,Temple23}.

\begin{acknowledgements}
  
We thank the referee for a constructive review of our work that contributed to improve our paper.
The paper is based on observations obtained with the {\sl XMM-Newton} ESA science
mission with instruments and contributions directly funded by ESA
member states and the USA (NASA). 
 This work made use of data from the
{\sl NuSTAR} mission, a project led by the California Institute of
Technology, managed by the Jet Propulsion Laboratory, and
funded by NASA. 
This research has made use of the {\sl NuSTAR} 
Data Analysis Software (NuSTARDAS) jointly developed by
the ASI Science Data Center and the California Institute of
Technology. This research has made use of NASA’s Astrophysics Data System.
This research has made use of the SIMBAD database,
operated at CDS, Strasbourg, France.
This research has made use of
the NASA/IPAC Extragalactic Database (NED) which is operated by the California Institute of Technology,
under contract with the National Aeronautics and Space Administration.
This work was supported by the French space agency (CNES). This research has made use
of computing facilities operated by CeSAM data centre at LAM,
Marseille, France. SH acknowledges support from the Science and Technology Facilities Council (STFC) through the studentship ST/V506643/1. CD acknowledges support from STFC through grant ST/T000244/1.
\end{acknowledgements}

%
%

\bibliographystyle{aa}
\bibliography{biblio}

\appendix

\section{Additional information about the X-ray broadband fits using the {\sc ReXcor} model}\label{sec:rexcorapp}

The {\sc reXcor} grid models do not include the underlying hard X-ray power-law continuum,  therefore we also add a cut-off power-law component by using {\sc zcutoffpl} \citep{Xiang22}. The high-energy cut-off values are fixed to the mean values found in \cite{Porquet21}  when fitting the data above 3\,keV with {\sc zcutoffpl}: 187\,keV in 2019 and 216\,keV. The weak Fe\,K$\alpha$ narrow core ($EW$$\sim$\,40\,eV,  \citealt{Porquet21}) is taken into account by including a Gaussian emission line ($\sigma$=10\,eV) at 6.4\,keV. We apply the model grids calculated for an Eddington ratio of 0.1. The baseline model is: {\sc tbnew(Gal)$\times$(reXcor+zcutoffpl+zgaussian)}. The 2019 and 2020 {\sl XMM-Newton} and {\sl NuSTAR} spectra are fit simultaneously with $h_{\rm f}$, $\Gamma$, $f_{\rm h}$, and $\tau_{\rm T}$ free to vary between the two epochs. For each epoch, the photon index of {\sc reXcor} and {\sc zcutoffpl} are tied together. \\

\begin{figure}[t!]
\begin{tabular}{c}
\includegraphics[width=0.9\columnwidth,angle=0]{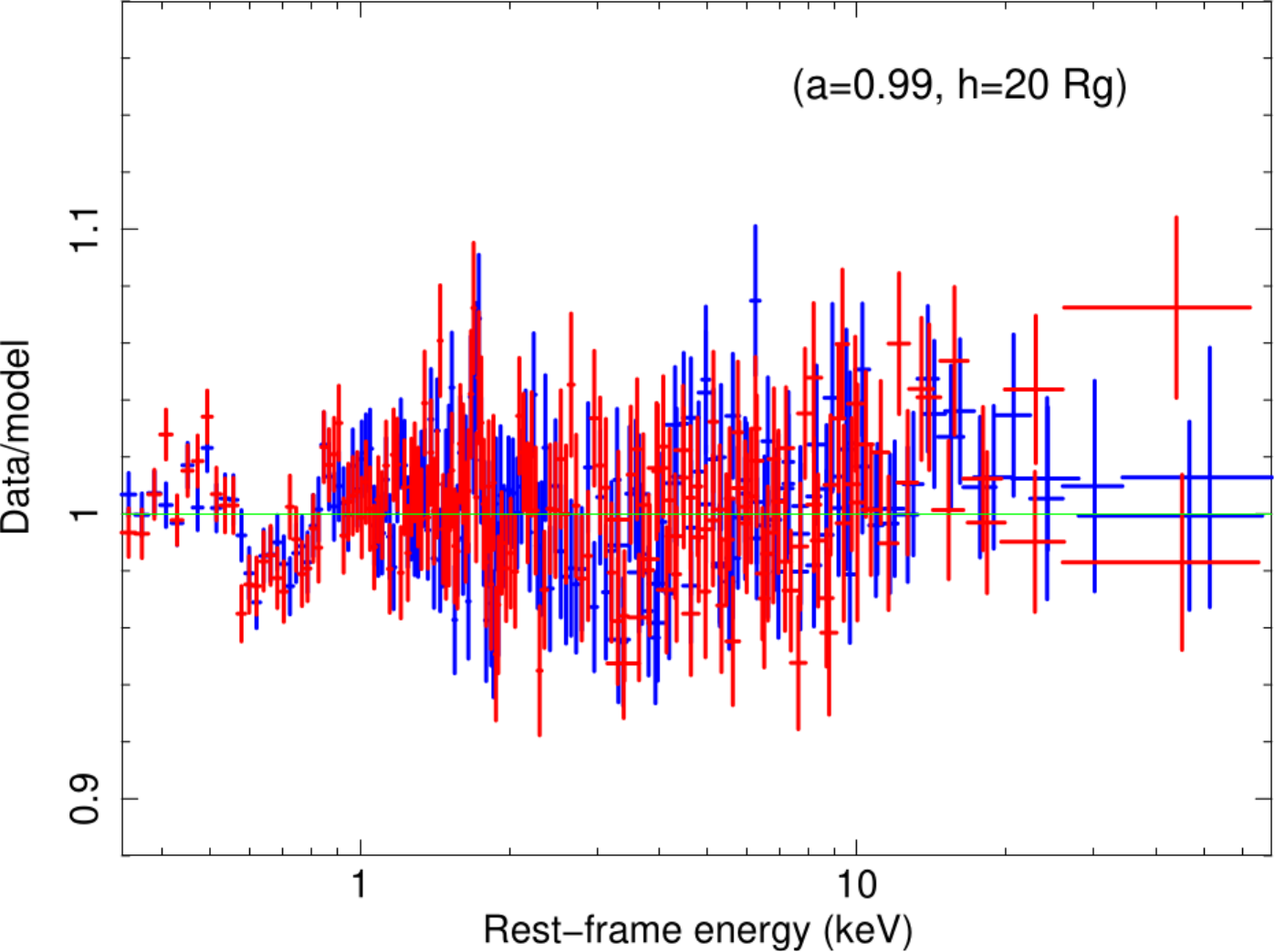} \\
\includegraphics[width=0.9\columnwidth,angle=0]{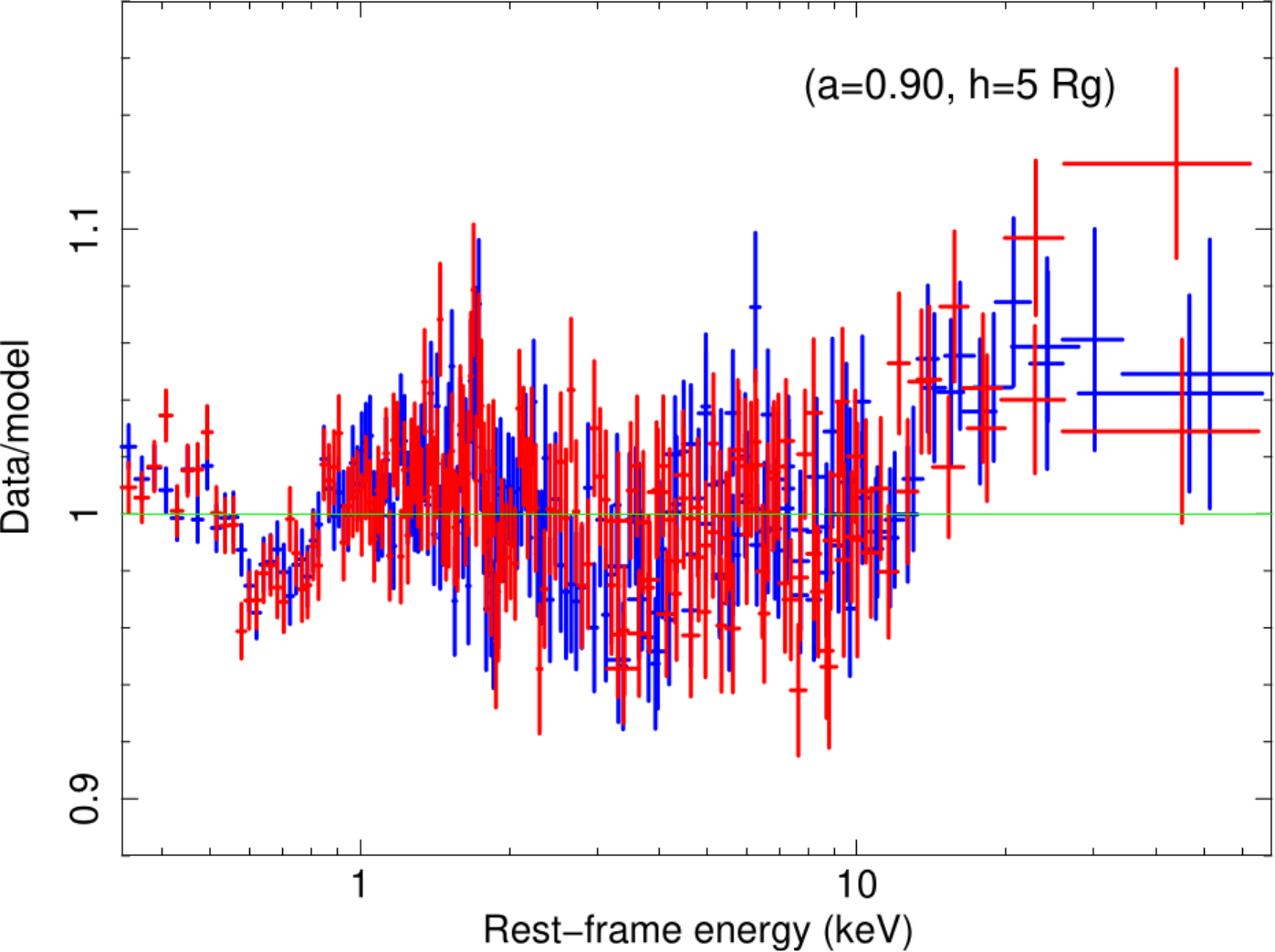} \\
\end{tabular}
\caption{Data-to-model ratio of the fits using the {\sc reXcor} model grids for the 2019 (blue) and 2020 (red) simultaneous {\sl XMM-Newton} and {\sl NuSTAR} spectra. The inferred parameter values are reported in Table~\ref{tab:rexcor}. Top panel: model calculated for a spin of 0.99 and a hot corona height of 20\,$R_{\sl g}$ ($\chi^{2}$/d.o.f.=1777/1605). Bottom panel: model calculated for a spin of 0.90 and a hot corona height of 5\,$R_{\sl g}$ ($\chi^{2}$/d.o.f.=1921/1605).}
\label{fig:rexcor}
\end{figure}

Statistically speaking, the best-fit result is found for the model grid calculated for a  spin value of 0.99 and a lamppost height of 20\,$R_{\rm g}$ (see Table~\ref{tab:rexcor} and Fig.~\ref{fig:rexcor}: top panel).  As shown in Fig.~\ref{fig:rexcor} (bottom panel), the data-model ratio for $a$=0.90 and $h$=5\,$R_{\rm g}$, which corresponds to the highest $\chi^2$ value, shows some noticeable deviation in the hard X-ray range. In all fits, there are small negative deviations of the data-model ratio of about 3\% at $\sim$0.5--0.6 keV, which is in the energy range of the \ion{O}{vii} triplet lines. As pointed out by \cite{Xiang22}, the sensitivity to temperature, density and optical depth of the He-like triplets \citep{Porquet10} are not correctly described by the {\sc reXcor} model and could lead to residual of a few percent. This slight discrepancy could also be due to the disc inclination of 30 degrees assumed when building the {\sc reXcor} grids, leading to  slightly larger line widths than for a system viewed almost face on; and/or to the oxygen abundance that could potentially be a bit overestimated, leading to an increase in the model of the \ion{O}{vii} line flux.

 \begin{table}[t!]
\caption{Best-fit results of the two simultaneous 2019 and 2020 X-ray broadband spectra ({\sl XMM-Newton} and {\sl NuSTAR})
 with a model of the form: {\sc tbnew(Gal)$\times$(reXcor + zcutoffpl + zgaussian)}.}
\centering
\begin{tabular}{@{}l c c}
\hline\hline
parameter & \multicolumn{1}{c}{2019 Nov}  & \multicolumn{1}{c}{2020 April}\\
\hline
 \multicolumn{3}{c}{(a=0.99, h=5)}\\
\hline
$f_X$  &  4.4$^{+0.6}_{-0.5}$$\times$10$^{-2}$ & 4.6$^{+0.8}_{-0.6}$$\times$10$^{-2}$ \\
$\Gamma_{\rm hot}$ & 1.78$\pm$0.01   & 1.73$\pm$0.01 \\
$h_f$    & 0.48$^{+0.01}_{-0.02}$  & 0.48$^{+0.02}_{-0.04}$  \\
$\tau_{\rm warm}$  & 14.7$^{+0.3}_{-1.5}$  & 13.0$^{+0.6}_{-1.1}$  \\
log\,F(reXcor)$^{(a)}$ &  $-$10.94$\pm$0.01 & $-$11.10$^{+0.01}_{-0.02}$    \\
log\,F(zcutoffpl)$^{(a)}$ &  $-$10.33$\pm$0.01 & $-$10.41$\pm$0.01 \\
$\chi^{2}$/d.o.f.\ ($\chi^{2}_{\rm red}$) & \multicolumn{2}{c}{1815.9/1605 (1.13)}  \\
\hline
 \multicolumn{3}{c}{(a=0.99, h=20)}\\
\hline
$f_X$    &  3.2$^{+1.2}_{-0.6}$$\times$10$^{-2}$ & 3.7$^{+1.1}_{-0.7}$$\times$10$^{-2}$ \\
$\Gamma_{\rm hot}$ & 1.79$\pm$0.01   & 1.75$\pm$0.01 \\
$h_f$    & 0.50$^{+0.03}_{-0.06}$  & 0.52$^{+0.02}_{-0.05}$  \\
$\tau_{\rm warm}$  & 15.5$\pm$1.5  & 15.4$^{+2.2}_{-1.0}$  \\
log\,F(reXcor)$^{(a)}$ &  $-$10.96$\pm$0.02 & $-$11.14$^{+0.02}_{-0.03}$    \\
log\,F(zcutoffpl)$^{(a)}$ &  $-$10.33$\pm$0.01 & $-$10.40$\pm$0.01 \\
$\chi^{2}$/d.o.f.\ ($\chi^{2}_{\rm red}$) & \multicolumn{2}{c}{1777.3/1605 (1.11)}  \\
\hline
\multicolumn{3}{c}{(a=0.90, h=5)}\\
\hline
$f_X$   &  6.0$^{+2.4}_{-1.2}$$\times$10$^{-2}$ & 6.3$^{+1.1}_{-0.9}$$\times$10$^{-2}$ \\
$\Gamma_{\rm hot}$  & 1.79$\pm$0.01   & 1.74$\pm$0.01 \\
$h_f$  & 0.62$^{+0.01}_{-0.02}$ & 0.62$^{+0.02}_{-0.01}$  \\
$\tau_{\rm warm}$  & 20.9$^{+1.9}_{-2.7}$  & 19.3$^{+0.6}_{-0.7}$  \\
log\,F(reXcor)$^{(a)}$ &  $-$10.98$\pm$0.01 & $-$11.15$\pm$0.01    \\
log\,F(zcutoffpl)$^{(a)}$ &  $-$10.33$\pm$0.01 & $-$10.39$\pm$0.01 \\
$\chi^{2}$/d.o.f.\ ($\chi^{2}_{\rm red}$) & \multicolumn{2}{c}{1921.3/1605 (1.20)}  \\
\hline
 \multicolumn{3}{c}{(a=0.90, h=20)}\\
\hline
$f_X$   &  4.0$^{+2.6}_{-1.2}$$\times$10$^{-2}$ & 5.5$^{+0.9}_{-1.3}$$\times$10$^{-2}$ \\
$\Gamma_{\rm hot}$ & 1.81$\pm$0.01  & 1.78$\pm$0.01 \\
$h_f$    & 0.63$^{+0.05}_{-0.02}$  & 0.72$\pm$0.02  \\
$\tau_{\rm warm}$  & 21.1$^{+4.8}_{-1.1}$  & 27.6$^{+1.5}_{-0.8}$  \\
log\,F(reXcor)$^{(a)}$ &  $-$11.00$\pm$0.01 & $-$11.23$\pm$0.01    \\
log\,F(zcutoffpl)$^{(a)}$ &  $-$10.32$\pm$0.01 & $-$10.39$\pm$0.01 \\
$\chi^{2}$/d.o.f. ($\chi^{2}_{\rm red}$) & \multicolumn{2}{c}{1834.0/1605 (1.14)}  \\
\hline    \hline
\end{tabular}
\label{tab:rexcor}
\flushleft
\small{\textit{Notes}. $^{(a)}$ 0.3--10\,keV pn unabsorbed fluxes (erg\,cm$^{-2}$\,s$^{-1}$).}
\end{table}

\section{Different SED models for the 2019 and 2020 simultaneous {\sl XMM-Newton} and {\sl NuSTAR} observations}\label{app:diffsed}
In table~\ref{tab:SEDfits}, we report the best-fit values found for different SED models for comparison purposes:
\begin{itemize}
 \item {\sc relagn+relxillcp}: same baseline as reported in $\S$\ref{sec:SED} (Table~\ref{tab:SED}):
{\sc relagn+relxillcp+zgaussian};
\item {\sc relagn+relxilllpcp}: same as the baseline model, except that a relativistic model assuming a lamppost geometry is used;
\item {\sc relagn}: no relativistic reflection component is included;
\item {\sc agnsed+relxillcp}: the {\sc agnsed} model is used instead of {\sc relagn}, that is to say no general relativistic ray-tracing is included;
\item {\sc agnsed}: same as the fourth row except that no relativistic reflection component is included.
\end{itemize}

The comparison of the best-fit values shows that similar results are found for {\sc relagn+relxillcp} and {\sc relagn+relxilllpcp}. Indeed, for Mrk\,110, the relativistic contribution is found to be weak \citep{Porquet21}, and so its exact modelling impact is negligible here. However, for the {\sc relagn} solo fit, the best-fit value is higher and the spin value is less contrained with a smaller lower limit leading to slightly higher $R_{\rm hot}$ and $R_{\rm warm}$ values (for a comparable $\dot{m}$ value).
Comparing {\sc relagn+relxillcp} and {\sc agnsed+relxillcp}, we find that the values of $\dot{m}$ and spin are significantly underestimated when the relativistic effects are not included. This confirmed the behaviour found by \cite{Hagen23a} for Fairall\,9, indeed, the `increase in Eddington ratio and spin is compensating for the reduction in observed power from the general relativistic raytracing'.

 \begin{table*}[h!]
\caption{Simultaneous SED fits
to the 2019 and 2020 simultaneous {\sl XMM-Newton}-pn and {\sl NuSTAR} spectra of Mrk\,110.
`(t)' means that the value has been tied between both epochs.
}
\centering
\begin{tabular}{@{}l c c c c c}
\hline\hline
parameter & {\sc relagn} & {\sc relagn} & {\sc relagn} & {\sc agnsed} & {\sc agnsed} \\
          & {\sc + relxillcp} & {\sc + relxilllpcp} &   & {\sc + relxillcp} &    \\
\hline
          \multicolumn{6}{c}{2019}\\
\hline
$a$                    & $\geq$0.997 &$\geq$0.996 & $\geq$0.82  & 0.84$^{+0.03}_{-0.17}$ & 0.81$^{+0.11}_{-0.23}$ \\
log\,$\dot{m}$              &  $-$1.03$^{+0.01}_{-0.03}$ & $-$1.04$^{+0.01}_{-0.03}$ & $-$1.06$^{+0.06}_{-0.19}$ & $-$1.49$\pm$0.01 & $-$1.46$\pm$0.02\\
$kT_{\rm hot}$ (keV)   & 58$^{+25}_{-8}$  & 55$^{+24}_{-6}$  &$\geq$76 &  34$^{+10}_{-6}$  & 47$^{+71}_{-15}$ \\
$\Gamma_{\rm hot}$ & 1.86$\pm$0.01 & 1.86$\pm$0.01  & 1.83$\pm$0.01 & 1.86$\pm$0.01 & 1.83$\pm$0.01 \\
$R_{\rm hot}$ (R$_{\rm g}$) & 16$^{+1}_{-4}$ & 16$^{+1}_{-5}$ & 18$^{+1}_{-2}$ & 14$^{+5}_{-3}$ & 15$^{+7}_{-4}$  \\
$kT_{\rm warm}$ & 0.23$\pm$0.01 & 0.23$\pm$0.01 & 0.23$\pm$0.01 & 0.22$\pm$0.01 & 0.22$\pm$0.01 \\
$\Gamma_{\rm warm}$ & 2.48$^{+0.02}_{-0.03}$  & 2.48$^{+0.02}_{-0.03}$ & 2.45$^{+0.02}_{-0.03}$ & 2.50$\pm$0.04 & 2.47$^{+0.04}_{-0.02}$\\
$R_{\rm warm}$  (R$_{\rm g}$)& 88$^{+4}_{-3}$ & 87$^{+3}_{-7}$ & 98$^{+12}_{-9}$ & 73$^{+29}_{-21}$ & 81$^{+36}_{-28}$ \\
log\,$\xi$ & 1.0$\pm$0.2 & 1.0$^{+0.1}_{-0.2}$ & $-$ & 1.0$^{+0.2}_{-0.1}$ & $-$\\
norm(relxill) & 1.1$\pm$0.2$\times$10$^{-5}$ & 3.6$^{+0.8}_{-0.4}\times$10$^{-5}$ & $-$ & 2.0$\pm$0.5$\times$10$^{-5}$ & $-$ \\
\hline
 \multicolumn{6}{c}{2020}\\
\hline
$a$ & (t) & (t) & (t) & (t) & (t)  \\
log\,$\dot{m}$  & $-$1.14$^{+0.01}_{-0.03}$ & $-$1.14$\pm$0.01  & $-$1.16$^{+0.02}_{-0.18}$ & $-$1.59$\pm$0.01 & $-$1.56$^{+0.03}_{-0.02}$ \\
$kT_{\rm hot}$ (keV)   & (t) & (t) & (t) & (t) & (t) \\
$\Gamma_{\rm hot}$ & 1.82$\pm$0.01 & 1.82$\pm$0.01 & 1.79$\pm$0.01 & 1.82$\pm$0.01 & 1.79$\pm$0.01  \\
$R_{\rm hot}$ (R$_{\rm g}$) & 20$\pm$1 & 20$\pm$1  & 23$\pm$1 & 18$^{+6}_{-3}$ & 19$^{+8}_{-5}$\\
$kT_{\rm warm}$ & 0.24$\pm$0.01 & 0.24$\pm$0.01 & 0.24$\pm$0.01 & 0.23$\pm$0.01 & 0.24$^{+0.02}_{-0.01}$   \\
$\Gamma_{\rm warm}$  & 2.46$^{+0.03}_{-0.04}$ & 2.45$\pm$0.03 & 2.42$^{+0.02}_{-0.03}$ & 2.47$\pm$0.04 & 2.44$\pm$0.04  \\
$R_{\rm warm}$  (R$_{\rm g}$) & 79$^{+3}_{-8}$ & 78$^{+3}_{-6}$ & 91$^{+10}_{-8}$ & 67$^{+18}_{-13}$& 76$^{+31}_{-23}$\\
log\,$\xi$ & (t) & (t) & $-$ & (t) & $-$ \\
norm(relxill) & 1.9$^{+0.4}_{-0.5}\times$10$^{-5}$ & 3.1$\pm$0.7$\times$10$^{-5}$ & $-$ & 1.7$\pm$0.5$\times$10$^{-5}$ &$-$\\
\hline
$\chi^{2}$/d.o.f. & 1725.0/1610 & 1726.3/1610 & 1793.2/1613 & 1725.4/1610 & 1789.1/1613  \\
$\chi^{2}_{\rm red}$ & 1.07 & 1.07 & 1.11 & 1.07 & 1.11 \\
\hline    \hline
\end{tabular}
\label{tab:SEDfits}
\end{table*}

\end{document}